\documentclass[aip, jap, showpacs, reprint, preprintnumbers,amsmath,amssymb,superscriptaddress,floatfix]{revtex4-1}

\usepackage[]{graphicx}      
\usepackage{bm}             	
\usepackage{times}			
\usepackage{tabularx}
\usepackage{color}
\usepackage{dcolumn}		
\usepackage{amsmath}
\usepackage{amssymb}
\usepackage{amsfonts}
\usepackage{times}
\usepackage{tabularx}
\usepackage{xcolor,colortbl}
\usepackage{ulem}
\usepackage{amsmath}

\makeatletter
\newcommand\xleftrightarrow[2][]{%
  \ext@arrow 9999{\longleftrightarrowfill@}{#1}{#2}}
\newcommand\longleftrightarrowfill@{%
  \arrowfill@\leftarrow\relbar\rightarrow}
\makeatother

\begin{document}

\title{\centering Exchange energies in CoFeB/Ru/CoFeB Synthetic Antiferromagnets}

\author{A. Mouhoub}
\email{asma.mouhoub@u-psud.fr}
\affiliation{Universit\'e Paris-Saclay, CNRS, Centre de Nanosciences et de Nanotechnologies, 91120 Palaiseau, France}
\author{F. Millo}
\author{C. Chappert}
\author{J.-V. Kim}
\author{J. L\'etang}
\affiliation{Universit\'e Paris-Saclay, CNRS, Centre de Nanosciences et de Nanotechnologies, 91120 Palaiseau, France}
\author{A. Solignac}
\affiliation{SPEC, CEA, CNRS, Université Paris-Saclay, 91191 Gif-sur-Yvette, France}
\author{T. Devolder}
\affiliation{Universit\'e Paris-Saclay, CNRS, Centre de Nanosciences et de Nanotechnologies, 91120 Palaiseau, France}

\date{\today}                                           
                       
%
%
\begin{abstract}
The interlayer exchange coupling confers specific properties to Synthetic Antiferromagnets that make them suitable for several applications of spintronics. The efficient use of this magnetic configuration requires an in-depth understanding of the magnetic properties and their correlation with the material structure. Here we establish a reliable procedure to quantify the interlayer exchange coupling and the intralayer exchange stiffness in synthetic antiferromagnets; we apply it to the ultrasmooth and amorphous Co$_{40}$Fe$_{40}$B$_{20}$ (5-40 nm)/Ru/ Co$_{40}$Fe$_{40}$B$_{20}$ material platform. The complex interplay between the two exchange interactions results in a gradient of the magnetization orientation across the thickness of the stack which alters the hysteresis and the spin wave eigenmodes of the stack in a non trivial way. We measured the field-dependence of the frequencies of the first four spin waves confined within the thickness of the stack. We modeled these frequencies and the corresponding thickness profiles of these spin waves using micromagnetic simulations. The comparison with the experimental results allows to deduce the magnetic parameters that best account for the sample behavior. 
The exchange stiffness is established to be 16 $\pm$ 2 pJ/m, independently of the  Co$_{40}$Fe$_{40}$B$_{20}$ thickness. The interlayer exchange coupling starts from -1.7 mJ/m$^2$ for the thinnest layers and it can be maintained above -1.3 mJ/m$^2$ for CoFeB layers as thick as 40 nm. The comparison of our method with earlier characterizations using the sole saturation fields argues for a need to revisit the tabulated values of interlayer exchange coupling in thick synthetic antiferromagnets.\\

\end{abstract}

\maketitle

%
%

\section{Introduction} 
Synthetic antiferromagnets (SAFs) are a class of artificial multilayers consisting of two identical ferromagnetic layers separated by a non-magnetic spacer that favors antiparallel magnetizations thanks to an interlayer coupling \cite{grunberg_layered_1986}. This interlayer coupling is an additional degree of freedom that confers a large tunability to SAFs, and allows the customization of their magnetic properties for specific applications. As stray-field-free magnets, SAFs triggered for instance interest as part of stable reference layers in sensors or as free layers in random access memory applications \cite{worledge_spin_2004, hayakawa_current-induced_2006}. SAFs have also been used in high performance spin-torque oscillators \cite{houssameddine_spin_2010}, or as a medium in which domain walls can reach exceptionally high velocities \cite{yang_domain-wall_2015}. Recently, SAFs entered the field of magnonics \cite{chumak_magnon_2015} where the remarkable anisotropy and non-reciprocity of their spin waves (SW) have attracted attention \cite{franco_enhancement_2020, gallardo_spin-wave_2021}.

It is therefore important to develop methods to properly measure the magnetic properties of SAFs and to understand their correlation with the material structure of the multilayer stack. The two layers of the SAF are coupled by two distinct phenomena: the (electron-mediated) interlayer exchange interaction \cite{bruno_theory_1995}, and the (roughness-mediated) so-called "N\'eel" dipolar coupling \cite{neel_sur_1962, schrag_neorange-peel_2000}. Their sum 
 is described by an interfacial energy $J_{\textrm{tot}}=J_\textrm{N\'eel}+J_\textrm{ex}$ or an equivalent field $H_{J}=-\frac{2 J_{\textrm{tot}}}{ {\mu_0 M_{s}}  {t_\textrm{mag}}}$, where $M_s$ is the saturation magnetization and $t_\textrm{mag}$ the thickness of each of the (identical) layers of the SAF; $J$ is conventionally deduced by confusing $H_J$ with the saturation field, i.e. the field that sets the two magnets of the SAF in a parallel state along the easiest direction \cite{nguyen_van_dau_magnetic_1988, parkin_oscillations_1990, wiese_magnetic_2005,  dai_strong_2021}. However, this approach is inaccurate when the magnetizations of the two layers are not strictly uniform across their thickness. This happens as soon as $t_\textrm{mag}$ becomes comparable or larger than one of the two characteristic lengths of the system: the bulk exchange length \cite{hubert_magnetic_2008} $\lambda_B=\sqrt{2A_\textrm{ex}/(\mu_0 M^2_s)}$, where $A_\textrm{ex}$ is the bulk (intralayer) exchange stiffness, and the depth $\lambda_I= A_\textrm{ex}/J_{\textrm{tot}}$ in which the magnetization orientation in the bulk of a sample feels the micromagnetic state at the two interfaces of the interlayer spacer.

In many of the currently used SAFs, the condition $t_\textrm{mag }\ll \{ \lambda_I,~\lambda_B \}$ is not fulfilled; we will for instance conclude that $\lambda_I \approx 10$ nm and $\lambda_B \approx 4$ nm in our SAFs. So
when a field is applied the magnetizations in the regions far from the spacer reorient, while those close to the interfaces with the spacer keep their magnetizations more antiparallel. The magnetic hysteresis and the spin waves are considerably modified by this gradient of the magnetization orientation. In this case, $J$ cannot be evaluated from the sole knowledge of the saturation field: a more elaborate method taking into account the competition between inter and intralayer exchange must be developed.

Here we study the material structure, the magnetic hysteresis, the spin wave frequencies and the spin wave thickness profile in a series of SAFs with relevant thicknesses spanning from near $\lambda_I$ and $\lambda_B$ to much thicker. By fitting the field dependence of the spin wave modes frequencies with thickness resolved micromagnetic simulations, we deduce the intralayer exchange interaction and interlayer coupling constants with quantifiable reliability. After correction from roughness effects, we observe that a strong interlayer exchange (electron mediated) \cite{bruno_theory_1995} interaction with $|J_\textrm{ex }| >$ 1.3 mJ/m$^2$ is maintained on structurally smooth SAFs for CoFeB layers as thick as 40 nm; this comes despite an apparently easy and very gradual saturation that arises from the gradient of the magnetization orientation which develops within the thickness of the stack.

\begin{figure} 
\hspace{-0cm}
\includegraphics[width=8.5cm]{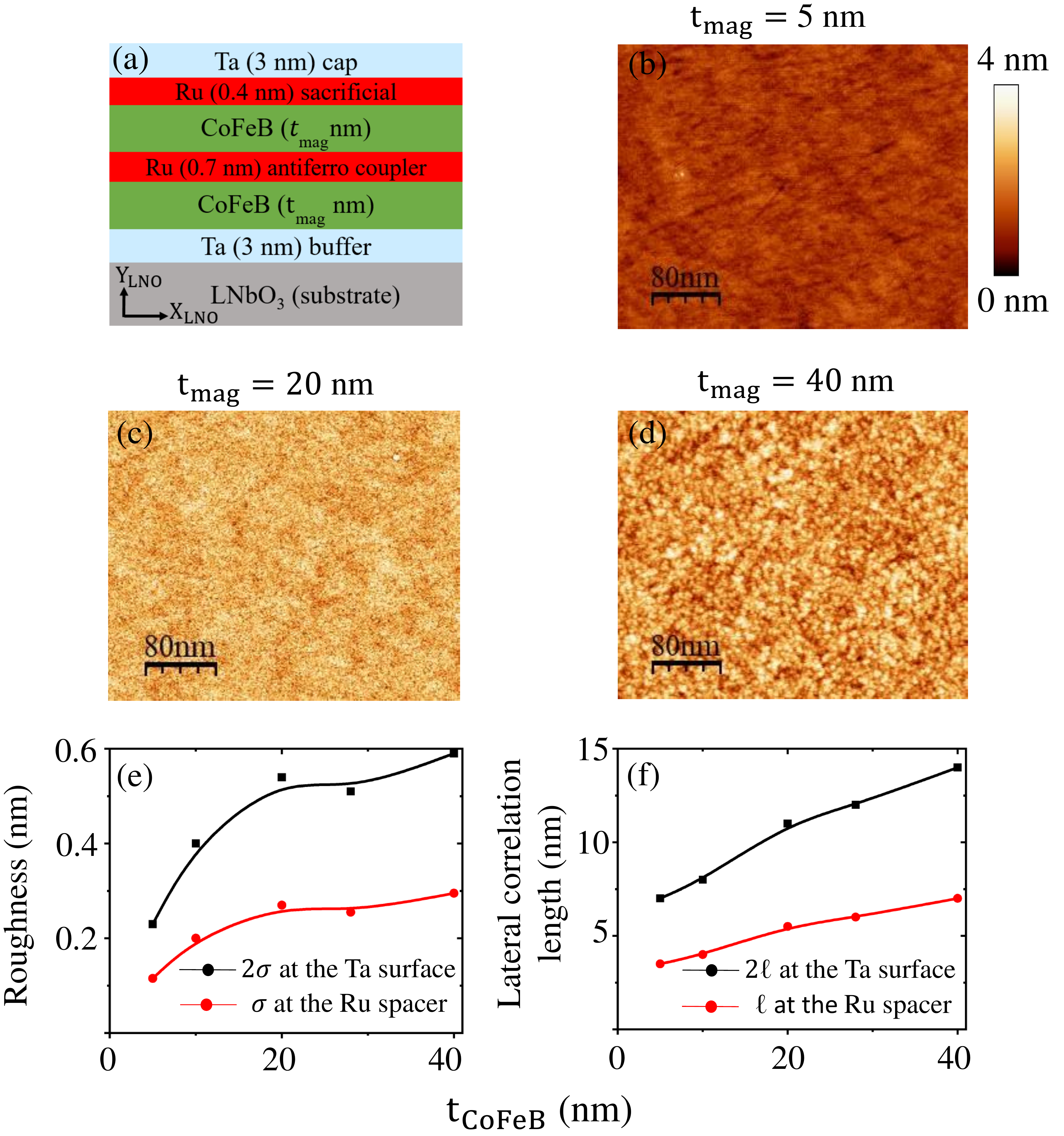}
\caption{Structural properties of the synthetic antiferromagnets. (a): Sketch of the nominal stack with the crystalline directions of the substrate. (b-d): AFM measurement of the surface topography ($1\times1~\mu\textrm{m}^2$) of the SAFs with $t_\textrm{mag}= 5,~20,~40~\textrm{nm}$. (e): Surface roughness of the SAF measured at the Ta surface (2$\sigma$) and interpolated at the Ru spacer ($\sigma$). (f): Lateral correlation length of the surface roughness  measured at the Ta outer surface (2$\ell$) and interpolated the buried Ru spacer ($\ell$).}
\label{figAFM}
\end{figure}

\section{Samples and methods \label{methods} \label{SectionSampleAndMethods}}
\subsection{Thin films growth and structure}
We grow our SAF multilayers by sputter-deposition on Y-cut LiNbO$_{3}$ (LNO) single crystal substrates in a chamber of base pressure below $10^{-7}$ mbar. The deposition was conducted under an (optimized) Argon pressure of $ 5 \times10^{-3}$ mbar, i.e. sufficiently low to maximize the magnetization \cite{cho_effects_2013}. 
The SAFs are symmetric with nominal composition LNO / Ta ( 6 nm)/ Co$_{40}$Fe$_{40}$B$_{20}$ ($t_\textrm{mag}$) / Ru (t$_\textrm{Ru}=7$ \r{A})/ Co$_{40}$Fe$_{40}$B$_{20}$ ($t_\textrm{mag}$) / Ru  (0.4 nm) / Ta (6 nm, cap), see Fig.~\ref{figAFM}(a). 
The investigated CoFeB thicknesses $t_\textrm{mag}$ are 5, 10, 15, 16.9, 20, 28 and 40 nm. The thickness of the Ru spacer was optimized to maximize the interlayer exchange coupling for our composition of CoFeB, in agreement with ref. \cite{wiese_antiferromagnetically_2004}. The Ru(0.4 nm) layer is a sacrificial layer that conveniently avoids \cite{swerts_beol_2015} the re-sputtering of the top CoFeB layer when the heavy and energetic Ta atoms of the cap impinge on the stack being grown.
Note that the biquadratic interlayer exchange coupling is known to be negligible for this $t_{Ru}$ and our CoFeB composition ratio \cite{hashimoto_fe_2006}.
The samples are in the as-grown state; $\theta-2\theta$ X-ray diffraction scans (not shown) argue for a bcc (011) and (112) texture of the Ta buffer layer and an amorphous state of the CoFeB layers, as anticipated for our Boron content \cite{kim_control_2022}.  

Additional samples containing a single CoFeB layer with a thickness of 17 nm were grown for the optimization of the deposition conditions. 
Vibrating Sample Magnetometry (VSM) and vector network analyzer ferromagnetic resonance (VNA-FMR, \cite{bilzer_vector_2007}) indicated that these reference samples have a saturation magnetization $\mu_0 M_s=$ 1.7 T and a damping $\alpha=0.0045 \pm 0.0005$. A tiny uniaxial anisotropy (an approximate field of 3 mT) was evidenced in the film plane; we will neglect it in the following.

Our present aim is to measure the interlayer exchange coupling \cite{bruno_theory_1995}. This requires to correct for the other source of interlayer coupling: the roughness-induced "orange-peel" coupling \cite{neel_sur_1962} that opposes the exchange coupling. The orange-peel coupling depends on the 
standard deviation $\sigma_\textrm{RMS}$ and the lateral wavelength $\ell$ of the conformal roughness of the Ru layer separating the two magnets \cite{schrag_neorange-peel_2000}. 
The values of $\ell$ and $\sigma_\textrm{RMS}$ at the buried Ru spacer layer are unfortunately not measurable by Atomic force microscopy (AFM); however they can be estimated from the interpolation of their values at the surface of the sample. AFM [Fig.~\ref{figAFM}(b-d)] indicates indeed that both the surface roughness and the lateral correlation length (the typical "grain size" at the top of the Ta cap layer) are quasi-linear functions of $t_\textrm{mag}$: thicker films have wider and taller grains (Fig.~\ref{figAFM}). We will thus assume that the values of $\ell$ and $\sigma_\textrm{RMS}$ at the Ru spacer layer are the halves of their (measured) value at the surface. The structure is clearly grainy [see eg. Fig.~\ref{figAFM}(d)]; however the autocorrelation function of the surface height measured by AFM has a single maximum (its $\delta$-correlation): it does not show any secondary peak that would indicate the existence of a most probable grain size from which an unquestionable value of $\ell$ could be given. We shall thus consider that the lateral wavelength $\ell$ of the roughness can be approximated by the full width at half maximum of this height autocorrelation function.  Note that a potential error in $\ell$ does not induce a large error on the N\'eel coupling since $J_\textrm{N\'eel}$ is only weakly dependent from $\ell$ when $\ell \gg t_{Ru} $. Indeed, following Ref.~\onlinecite{schrag_neorange-peel_2000}, the orange peel coupling energy is: \begin{equation} J_\textrm{N\'eel}=  2 \sqrt{2} {\pi^2}  \frac{(\sigma_\textrm{RMS})^2}{\ell } \mu_0 M_s^2 \exp{(-2 \pi \sqrt{2} \frac{t_{Ru}}{ \ell})},\label{neel} \end{equation}which can be evaluated using the topographical data of Fig.~\ref{figAFM}(e, f).
Note that the above equation assumes a 1-dimensional periodic roughness \cite{neel_sur_1962}, which in principle cannot be applied to any topography with a broad 2-dimensional distribution of the in-plane periods $\ell$. Eq. \ref{neel} can however be reliably used because its variation is dominated by that of $\sigma_\textrm{RMS}$. 
 The orange-peel coupling is negligible for our thinnest layers but that it should rise to +0.33 $\pm 0.02$ mJ/m$^2$ 
for $t_\textrm{mag} \geq 20~\textrm{nm}$ when a grainy topography becomes perceivable [see Fig.~\ref{figAFM}(c)]. These values and their uncertainty will be used to identify the pure exchange part in the \textit{total} interlayer coupling energies that we will determine in the results section.

\subsection{Magnetic measurements}

\subsubsection{Magnetic hysteresis}
\begin{figure} 
\hspace{-0.2cm}
\centering
\includegraphics[width=7.5cm]{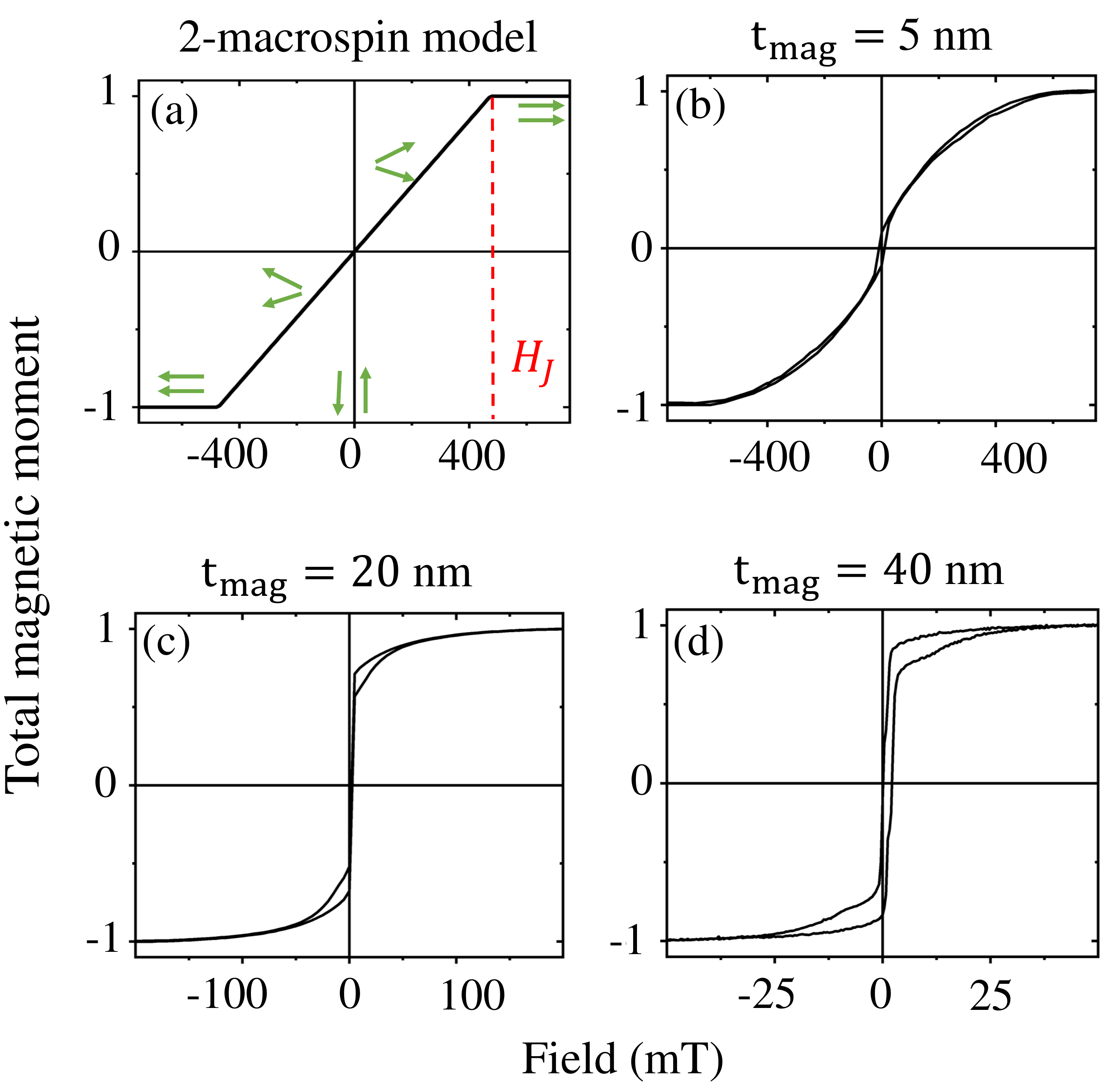}
\caption{Quasi-static magnetic properties of the synthetic antiferromagnets. (a): Simulated hysteresis loop of the 2-macrospin model. (b-d): Experimental hysteresis loops of the SAFs for $t_\textrm{mag}= 5,~20,~40 ~\textrm{nm}$.} 
\label{figVSM}
\end{figure}

The hysteresis loops of the films measured by VSM are reported in Fig.~\ref{figVSM}(b-d). They will be fitted to their theoretical counterparts obtained from exact micromagnetic simulations (Fig.~\ref{figureModeledLoops}); however for a start it is convenient to compare them to a toy model in which two identical macrospins (labeled 1 and 2) are coupled though a total interlayer coupling $J_\textrm{tot}= J_\textrm{ex}+J_\textrm{N\'eel}$. In this 2-macrospin description the loop is linear [Fig.~\ref{figVSM}(a)], i.e. $\frac{M_x}{M_S}=\frac{H_x}{H_{J,\textrm{tot}}}$ until a clear saturation for $|H_x| \geq H_{J,\textrm{tot}}$: The AntiParallel (AP) remanent state evolves to the saturated (parallel) state through a gradual scissoring of the macrospins (see sketches in [Fig.~\ref{figVSM}(a)]). 

While very thin SAFs can have quasi-linear hysteresis loops resembling the 2-macrospin approximation (see for instance refs.\cite{wiese_magnetic_2005, devolder_spin_2012}), the loop of our thinnest SAF sample ($t_\textrm{mag} = 5~\textrm{nm}$) already shows a clear rounding [Fig.~\ref{figVSM}(b)]. The saturation in thicker SAFs requires smaller fields [Fig.~\ref{figVSM}(c, d)], reflecting the thickness dependence of the exchange field. Unfortunately, the rounding of the loop becomes too pronounced to define a saturation field in the presence of experimental noise. This rounding will be discussed later together with the dynamical properties of the SAFs. A slight opening of the hysteresis loop is present in the thickest films [Fig.~\ref{figVSM}(d)]; there the structural defects (e.g. the grainy structure) and the small but non-vanishing anisotropy probably leads to some irreversibility in the magnetization process. 


\subsubsection{Spin waves}
\begin{figure}
\hspace{-0.2cm}
\includegraphics[width=8.6cm]{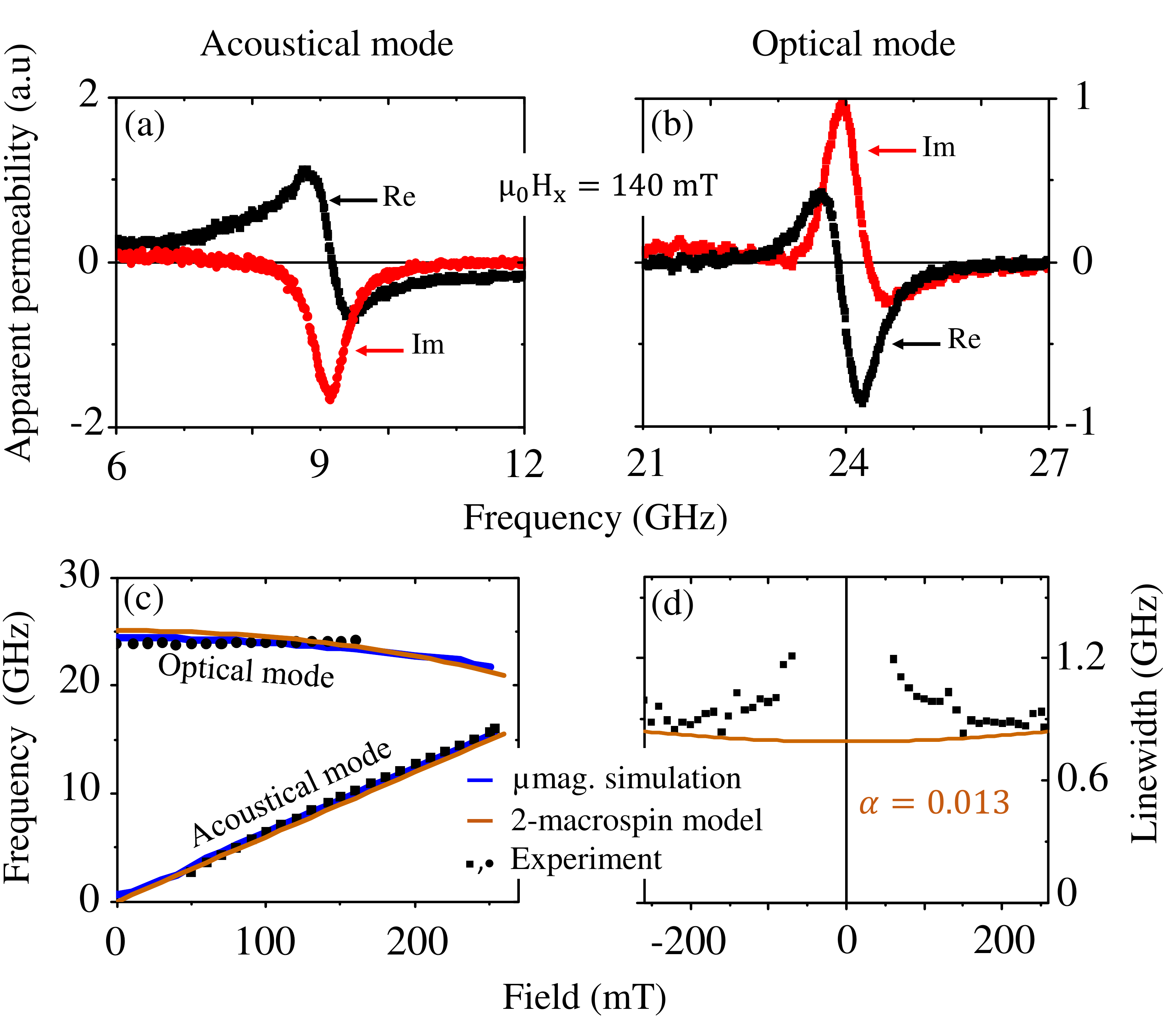}
\caption{Dynamic magnetic properties of a SAF with $t_\textrm{mag}=5~\textrm{nm}$.
Apparent permeability for an applied field of 140 mT in frequency windows around the acoustical (panel a) and optical (panel b) spin wave modes. (c) Field dependence of the mode frequencies, compared to the 2-macropin model and to the micromagnetic model with a thickness stratification $n_z=16$. The models are evaluated with 
$M_s=1.35\times 10^6 ~\textrm{A/m}$, $J_\textrm{tot}$=-1.64 ~mJ/m$^2$ and $A_\textrm{ex}=16 ~\textrm{pJ/m}$. (d) Field-dependence of the linewidth (FWHM) of the acoustical mode, and comparison with the Gilbert linewidth $ \frac{\Delta\omega_\textrm{ac}}{2 \pi} =\frac {\alpha \gamma_0} {2 \pi} (M_s + {\frac{H_J^2+H_x^2}{H_J}})$ for a damping of $\alpha=0.013$.}
\label{figureVNAFMR}
\end{figure}

The frequencies $f_{j, \textrm{exp}}$ of the spin waves (SW) of the samples were identified using Vector Network Analyzer FerroMagnetic Resonance (VNA-FMR~\cite{bilzer_vector_2007}) in in-plane applied fields up to $\mu_0 H_x=250~\textrm{mT}$. In this method, the SAF is inductively coupled to a microwave coplanar waveguide. The frequency dependence of the impedance of this ensemble is used to extract the transverse susceptibility spectrum of the sample. Fig.~\ref{figureVNAFMR}(a, b) display representative spectra for both the acoustical and optical spin wave modes \cite{stamps_spin_1994} of the stack. Owing to the metallic nature of the SAF, the electromagnetic fields of the coplanar waveguide are shielded in a frequency-dependent manner \cite{bailleul_shielding_2013}, such that the apparent susceptibility can be rotated in the complex plane. To correct this, the spin wave modes are simply identified as maxima in the modulus of the apparent susceptibility, and the linewidth of each mode is set by a generalized Lorentzian fit. For the thinnest SAFs $t_\textrm{mag} \leq 20 ~\textrm{nm}$, only two spin wave modes could be detected [see the example in Fig.~\ref{figureVNAFMR} for $t_\textrm{mag} = 5~\textrm{nm}$]. Four SW modes were detected for the thicker SAFs as will be further discussed in Fig.~\ref{FversusHandModeProfiles}(a).

\subsection{Magnetic models} 
 
\subsubsection{Spin waves in the 2-macrospin model}
For an approximate description of the thinnest SAFs we first stick to the 2-macrospin model. For $0 \leq H_x < H_{J,\textrm{tot}}$ the system is in the scissor state and possesses two excitation modes: the acoustical and optical spin wave modes of frequencies\cite{devolder_measuring_2022} $
\frac   {\omega_\textrm{ac}} {2 \pi}= \frac{\gamma_0} {2 \pi} H_x \sqrt{\frac{M_s+H_{J,\textrm{tot}}}{H_{J,\textrm{tot}}}}
$ and $ \frac{\omega_\textrm{op}}{2 \pi} = \frac{\gamma_0}{2 \pi} \sqrt{\frac{M_s}{H_{J,\textrm{tot}}}} \sqrt{H_{J,\textrm{tot}}^2-H_x^2}$ where $\gamma_0$ is the gyromagnetic ratio. This simplified description seems still valid for our thinnest SAF with $t_\textrm{mag}=5~\textrm{nm}$, as plotted in Fig.~\ref{figureVNAFMR}(c).

\begin{figure*} 
\includegraphics[width=18cm \hspace{-0.4 cm}] {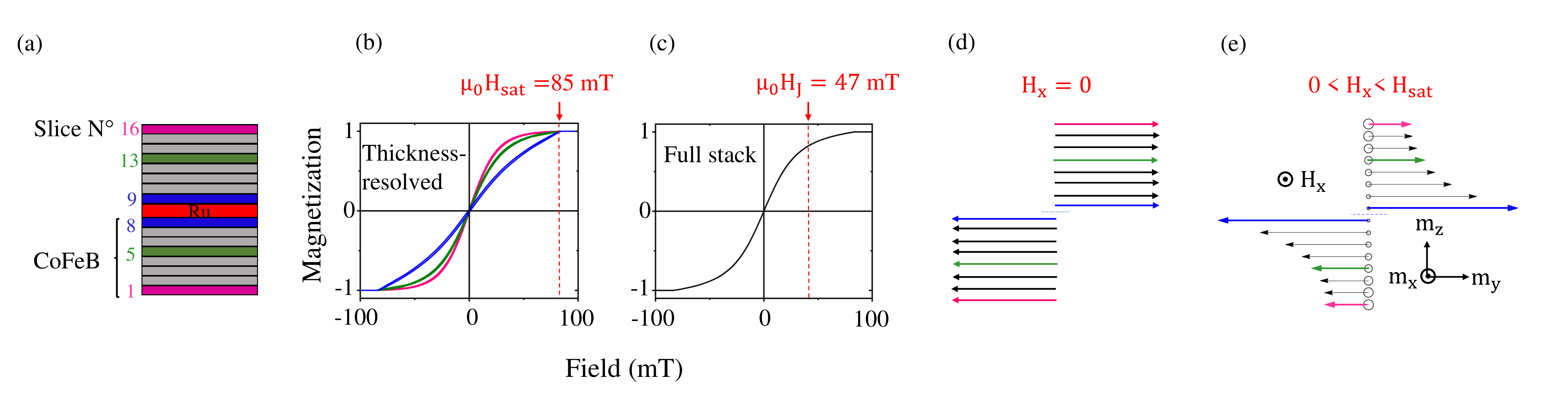}
\caption{Thickness-resolved hysteresis loop of a SAF with $t_\textrm{mag}=40~\textrm{nm}$ from micromagnetic simulations. (a) Sketch of the chosen stratification of the CoFeB layers. (b) Selected slice-resolved hysteresis loops and (c) total moment of the stack. (d) and (e): Sketch of the gradient of magnetization orientation along z at remanence and at finite applied field. (b) and (c) are evaluated for $t_\textrm{mag}=40~\textrm{nm}$, $M_s=1.35\times 10^6 ~\textrm{A/m}$, $J_\textrm{tot}=J_\textrm{ex}$=~-1 mJ/m$^2$ and $A_\textrm{ex}$=16 \textrm{pJ/m}. The arrows in (b) and (c) recall the values of the saturation field $H_\textrm{sat}$ and the interlayer exchange field $H_J$. }
\label{figureModeledLoops}
\end{figure*}

For thicker SAFs, the magnetizations within the thickness of each CoFeB layer can twist at a moderate exchange energy cost. In this case we need a full micromagnetic description for which we use the mumax3 software \cite{vansteenkiste_design_2014}. 

\subsubsection{Hysteresis in the micromagnetic description}
In our micromagnetic simulations, the material is described with a saturation magnetization $\mu_0 M_s=1.7~\textrm{T}$ and a damping $\alpha$=0.0045. The simulated shape is a cuboid of dimensions $\{320~\textrm{nm} \times 320 ~\textrm{nm} \times 2 t_\textrm{mag}\}$. Periodic boundary conditions are set in the sample plane (i.e., $xy$) to mimic an infinite thin film. The sample is meshed into $128\times 128\times n_z$ cells. The stratification $n_z=16$ is chosen in the thickness direction to have cells thinner than $\lambda_{B}$ even for our thickest SAFs. We disregard the spacer thickness and implement a direct interlayer coupling between the two magnetic layers of the SAF, i.e. at the frontier between the slices $i=8$ and 9. For each applied field $H_x$, we let the system relax to its ground state, allowing to plot either the hysteresis loop of the total moment $\frac{1}{n_z} \sum_{i=1}^{n_z} \frac{M_x^i}{M_s}$ versus the applied field $H_x$ [Fig.~\ref{figureModeledLoops}(c)], or the hysteresis loop of a specific slice $M_x^{i}/{M_s} $. The Fig.~\ref{figureModeledLoops}(b) shows for instance the loop of the Ru-facing slice ($i=8$, blue), and the top slice ($i=1~\textrm{or}~16$, magenta). The corresponding gradient of the magnetization orientation across the thickness is sketched in  [Fig.~\ref{figureModeledLoops}(e)].

\subsubsection{Micromagnetic determination of the spin waves}
We now aim to study the laterally uniform spin waves (SW) of the system in a thickness-resolved manner. To numerically excite these SWs we let mumax3 calculate the slice-resolved response $\vec{m}_{i}(t)$ of the system to a laterally uniform out-of-plane field pulse superimposed on the static applied field $H_x\vec{e}_x$. In the slice $i$ the pulse reads: $$\vec{h}^\textrm{pulse}(i,t > 0)=  \mathcal{H}(4-i) \frac{\sin(\pi f_c t)}{\pi f_c t}h \vec{e}_z,$$where $f_c=100~\textrm{GHz}$ and $h=0.1$ mT. The heaviside function $\mathcal{H}$ is to apply the field pulse in the sole bottom quarter of the SAF (slices $1 \leq i \leq 4$) in order to excite all modes including the ones that are non-uniform across the thickness. We let the simulation run until t$_\textrm{simu}=5$ ns.

We identify the SW frequencies as those at which the power spectra $|| \tilde{\vec M}_{i,~x}(f) ||^2$ are maximal for arbitrarily chosen $\tilde{\vec M}(f)$ component (here $M_x$) and slice. The "  $\,\tilde{ }\,$  " symbol recalls the complex-valued nature of $\tilde{\vec M}$. The frequency-domain magnetization $\tilde{\vec M}(f)$ is the Fourier transform of the Hann-window apodized version of $\vec{m}_{i}(t)$:
\begin{equation}
\tilde{\vec M}_{i}(f)= \mathcal{F}\left(\cos(\frac{\pi (t-t_\textrm{simu}/2)}{t_\textrm{simu}}) \, \vec{m}_{i}(t)\right)
\end{equation}

\begin{figure*} 
\includegraphics[width=18 cm]{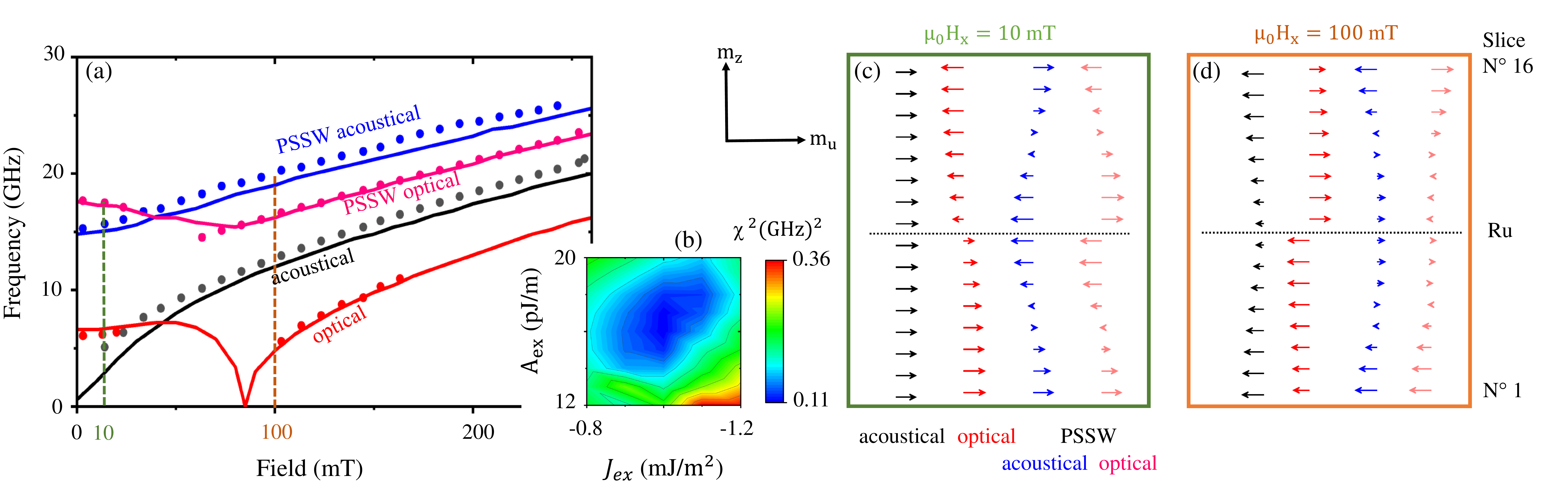}
\caption{Spin waves of a SAF with $t_\textrm{mag}=40~\textrm{nm}$. (a) Experimental frequencies of the spin wave modes (symbols) versus micromagnetic modeling (solid lines). (b): Map of the distance $\chi_\textrm{fit}^2$ between the experimental set of SW frequencies and the model in the $\{A_\textrm{ex}, J_\textrm{tot}\}$ plane. (c) Simulated thickness profiles of the dynamic magnetic components for the four spin wave modes at an applied field of $\mu_0 H_{x}$ = 10 mT. The dynamical component are plotted in the plane transverse to the magnetization ground state. (d) Ibidem at a larger applied field of $\mu_0 H_{x}$ = 100 mT. The micromagnetic modeling is implemented with $t_\textrm{mag}=40~\textrm{nm}$, $M_s=1.35\times 10^6 ~\textrm{A/m}$, $J_\textrm{tot}=J_\textrm{ex}$=~-1 mJ/m$^2$ and $A_\textrm{ex}$=16 \textrm{pJ/m}.}
\label{FversusHandModeProfiles}
\end{figure*}

We restrict our analysis to the $j=1,.. 4$ lowest frequency modes. Fig.~\ref{FversusHandModeProfiles}(a) shows the field dependence of their calculated frequencies $f_j(H_x)$ for a set of material parameters. To ease the discussion we shall label the SW modes according to the thickness profile of their dynamic magnetization $\mathcal{R}e \left( {\tilde{\vec M}}_{i}(f_j) \right)$ [see Fig.~\ref{FversusHandModeProfiles}(b, c)]. Conventionally, acoustic (respectively optical) modes have responses that are in-phase (resp. opposite-of-phase) on either sides of the spacer when displayed in the precession planes $\{\vec M_i \times \vec e_z, \vec e_z\}$ transverse to the ground state. Besides, the modes with amplitude nodes within the interior of each of the two magnetic layers recall the perpendicular standing spin waves (PSSW) of single layers \cite{bayer_spin-wave_2005}. Looking at the low-field mode profiles in Fig.~\ref{FversusHandModeProfiles}(b) we shall refer to the $j=1$ to 4 modes as the acoustical FMR, the optical FMR, the acoustical PSSW and the optical PSSW.

\subsubsection{Determination of the best-fitting material parameters}
We target to determine the values of $A_\textrm{ex}$ and $J_\textrm{tot}$ with the best possible accuracy. We thus assess the adequacy of any chosen set of material parameters by calculating the distance $\chi_\textrm{fit}$ between the experimental SW frequencies and their simulated counterpart. This distance is defined as: \begin{equation}
\chi_\textrm{fit}^2={\frac{1}{N} \sum_{j=1}^{N} (f_{j (H_x),~\textrm{exp}}- f_{j (H_x),~\textrm{sim}})^2},
\end{equation}
where $N=2$ for $t_\textrm{mag} \leq 20~\textrm{nm}$ (and $N=4$ otherwise) is the number of experimentally perceived SW modes. 
The best-fitting values of the material parameters for each SAF are found by minimizing $\chi_\textrm{fit}$ in the $\{A_\textrm{ex}, ~J_\textrm{tot}\}$ plane, as illustrated for the SAF with $t_\textrm{mag}=40~\textrm{nm}$ in the inset of Fig.~\ref{FversusHandModeProfiles}(a). We define the uncertainty on $\{A_\textrm{ex}, ~J_\textrm{tot}\}$ as the domain in which the distance $\chi_\textrm{fit}$ stays below twice its optimal value. The optimal values and their uncertainty are reported in Table I.

\section{ResultS}


The aim of this work is to measure the interlayer and intralayer exchange energies. 
As a first remark, we want to emphasize that although this was often practiced in the literature, this cannot be done from the sole hysteresis loop and the 2-macrospin model. Indeed as illustrated in Fig.~\ref{figureModeledLoops}(b-c) for $t_\textrm{mag}=40~\textrm{nm}$, confusing the saturation field $\mu_0 H_\textrm{sat} = 85~\textrm{mT}$ with the interlayer exchange field $\mu_0 H_J = 47~\textrm{mT}$ would lead to dramatically overestimate $J$. The exact same conclusion would be drawn if $H_\textrm{sat}$ was instead determined from the softening of the optical mode [see Fig.~\ref{FversusHandModeProfiles}(a)].

Alternatively, one could try and fit the experimental loops with the micromagnetically modeled ones, as practiced in ref. \cite{eyrich_effects_2014}. Unfortunately the unavoidable noise and the substrate-induced parasitic slope in the experimental loops make it difficult to determine the true saturation field, especially when a rounding is present at the saturation. On the contrary, the frequencies of the SAF eigenmodes can be determined with a high degree of confidence from both the experimental data \textit{and} the micromagnetic simulations.

We thus deduced the micromagnetic parameters $\{A_\textrm{ex}, ~J_\textrm{tot}\}$ of the SAF from the sole criterion of matching the experimental and simulated SW frequencies through a minimization of $\chi_\textrm{fit}$. The obtained best-fitting values $J_\textrm{tot}$  represent the \textit{total} interlayer coupling, i.e. accounting for the sum $J_\textrm{ex}+J_\textrm{N\'eel}$ of the interlayer exchange term and the orange-peel term. Table I gathers the best-fitting values of $\{A_\textrm{ex}, ~J_\textrm{tot}\}$ and their confidence interval. Note that the exchange stiffness could not be determined for $t_\textrm{mag}=5~\textrm{nm}$. For the thinnest films, the PSSW modes were not detected experimentally and the frequencies of the acoustic and optical SW were almost insensitive to $\{A_\textrm{ex}, ~J_\textrm{tot}\}$ in the simulations. As expected, the 2-macrospin model and the micromagnetic simulations match for this specific thickness only.

\begin{table}[!h] 
\caption{Micromagnetic parameters that best describe the spin wave frequencies and linewidth of the  Co$_{40}$Fe$_{40}$B$_{20}$ / Ru (7~\r{A}) /  Co$_{40}$Fe$_{40}$B$_{20}$ synthetic antiferromagnets. The symbol $\ast$ indicates that $\ell$ and $\sigma_\textrm{RMS}$ were extrapolated instead of measured.}
\label{FeCoB}
\begin{tabular}{| c | c | c | c |}
  \hline
Sample  & Exchange stiffness  & Interlayer coupling & Exchange part  \\ 
$t_\textrm{mag}$ (nm)  & $A_\textrm{ex}$ (pJ/m)  & $J_\textrm{ex}+J_\textrm{N\'eel}$ (mJ/m$^2$) & $J_\textrm{ex}$ (mJ/m$^2$)  \\ 
\hline
5 nm   & $\textrm{Unmeasurable}$  & -1.64 $\pm $ 0.14 & -1.68 $\pm $ 0.16    \\ \hline 
10 nm   & 13.4 $\pm  $ 5 & -1.71 $\pm$ 0.15 & -1.85  $\pm $ 0.19 \\ \hline
15 nm   & 16.2 $\pm $ 2.9  & -1.7 $\pm$ 0.1 & -1.9 $\ast$  \\ \hline 
17 nm   & 16.2 $\pm $ 3.3   & -1.71 $\pm$ 0.1 & -1.94 $\ast$ \\ \hline 
20 nm   & 14.8 $\pm $ 1.6  & -1.51 $\pm$ 0.1 & -1.78 $\pm $ 0.12 \\ \hline 
28 nm   & 16.4 $\pm $ 1.8 & -1.08 $\pm$ 0.14 & -1.33$\pm $  0.15 \\ \hline 
40 nm   & 16.4 $\pm $ 2.5 & -1 $\pm$ 0.15 & -1.33  $\pm $  0.17\\ \hline

\end{tabular}
\end{table}

Table I indicate that the intralayer exchange stiffness of amorphous Co$_{40}$Fe$_{40}$B$_{20}$ is 16 $\pm 2$ pJ/m and seems independent of the film thickness within our measurement accuracy. The interlayer exchange coupling is -1.7$\pm 0.15$ mJ/m$^2$ for the thinnest films and can be maintained above -1.3$\pm 0.15$ mJ/m$^2$ on structurally smooth SAFs for CoFeB layers as thick as 40 nm.

\begin{figure}[h] 
\hspace*{-0.5cm}
\includegraphics[width=9cm]{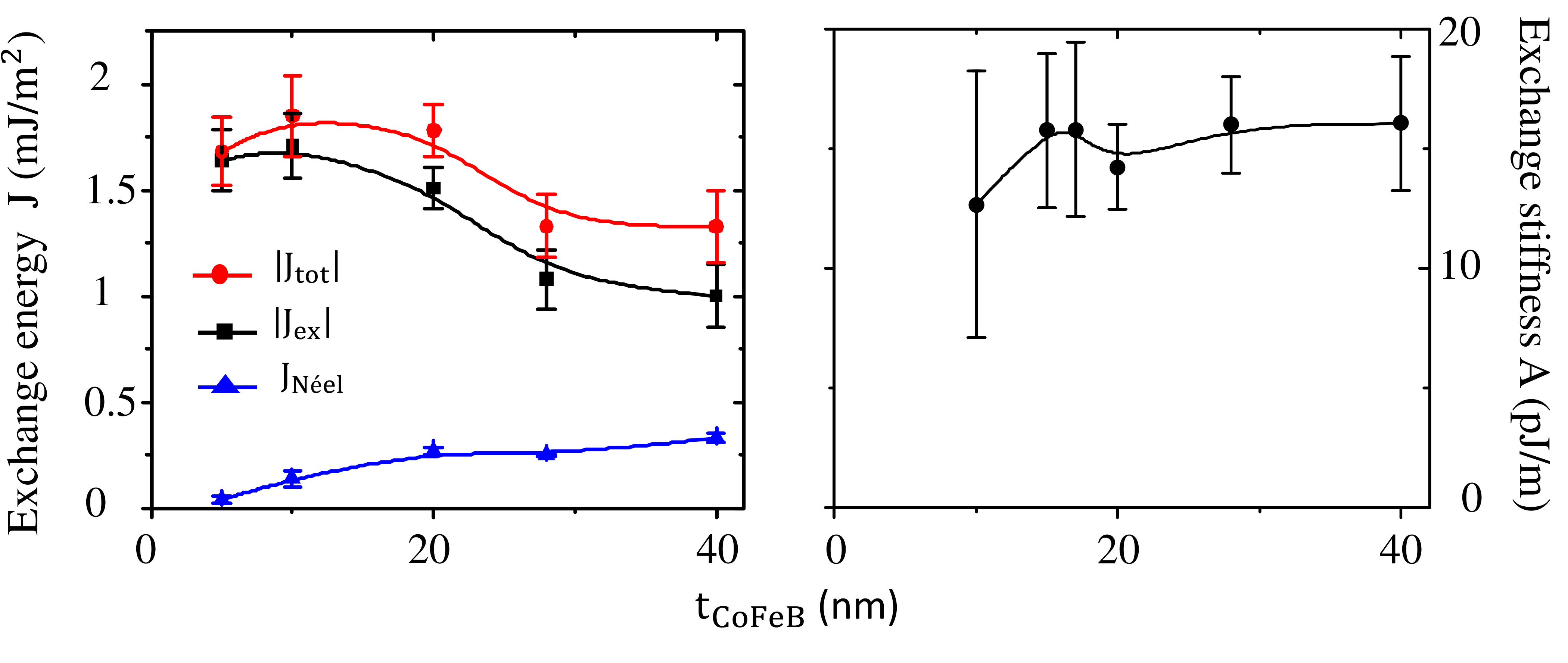}
\caption{Co$_{40}$Fe$_{40}$B$_{20}$ thickness dependence of (a): the interlayer coupling energies and (b): the intralayer exchange stiffness.}
\label{Exchange energiers J&A}
\end{figure}
\section{Discussion}


\subsection{Exchange stiffness}

Let us first discuss the value of the exchange stiffness, found to be 
$A_\textrm{ex}=16\pm 2~\textrm{pJ/m}$ for our amorphous Co$_{40}$Fe$_{40}$B$_{20}$ layers. This consolidated value is comparable to earlier reports for \textit{amorphous} alloys of composition equal to ours: Values of 11, 13 and 14 pJ/m were found respectively in the references \cite{cho_effect_2015, choi_exchange_2020, cho_effects_2013}. The exact value of $A_\textrm{ex}$ was also reported to depend on the argon pressure used for the sputter-deposition \cite{cho_effects_2013}: values were scattered from 10 to 14 pJ/m, the latter value being obtained for an argon pressure equal to ours. 

Despite this reasonable agreement with earlier reports, our number of $16\pm 2~\textrm{pJ/m}$ may be regarded as surprisingly small compared to that of \textit{crystalline} Co-Fe binary alloys such as elemental iron (20-23 pJ/m, \cite{cochran_light_1994, devolder_compositional_2013}), elemental cobalt (15.2, 28.5, 30 pJ/m, in \cite{eyrich_effects_2014, hillebrands_ultrathin_1994, choi_exchange_2020}]), Co$_{90}$Fe$_{10}$ (29.8 pJ/m in \cite{cho_effects_2013} and 25 pJ/m in \cite{chen_designing_2015}) or Co$_{80}$Fe$_{20}$ (26.1 pJ/m in \cite{bilzer_study_2006}). 
To understand this quantitative difference, it is worth keeping in mind that for a given composition CoFeB alloys have substantially smaller $A_\textrm{ex}$ when in the amorphous state as compared to when in the crystalline state \cite{cho_effect_2015, kim_control_2022}. In a comprehensive study  \cite{kim_control_2022}, J.-S. Kim et al. showed for instance that $A_\textrm{ex}$ drops from 16 to 8 pJ/m when rendering amorphous an alloy films of approximate composition Co$_{9}$Fe$_{86}$B$_{5}$. Note that these values are small because they refer to alloys on the Fe-rich side of the Co-Fe binary system.

Our conclusion that the most probable value of the exchange stiffness is $A_\textrm{ex}=16 \pm 2~\textrm{pJ/m}$ for amorphous Co$_{40}$Fe$_{40}$B$_{20}$ layers, therefore, calls for two main comments: it highlights the importance of (i) adjusting the Fe-Co composition and of (ii) controlling its crystalline or amorphous state.

\subsection{Interlayer exchange coupling}
Let us now comment on our estimate of the interlayer exchange coupling $J_\textrm{ex}=-1.7\pm0.15 ~\textrm{mJ/m}^2$. The comparison with literature is more problematic as many references (e.g. \cite{wiese_antiferromagnetically_2004}) either omit to report structural data (thus confusing potentially $J_\textrm{ex}$ and $J_\textrm{tot}$) and/or deduce their interlayer exchange coupling from a confusion between $H_\textrm{sat}$ and $H_J$. As already highlighted in Fig.~\ref{figureModeledLoops}, the confusion between $H_\textrm{sat}$ and $H_J$ leads to an overestimation of $J_\textrm{ex}$ for films thicker than $\lambda_I$ and $\lambda_B$. In contrast, the confusion between $J_\textrm{ex}$ and $J_\textrm{tot}$ leads to an underestimation of $J_\textrm{ex}$, but only for non-smooth stacks. We thus emphasize that a thorough analysis is required when looking back at literature data on the interlayer exchange coupling.

In the study of Waring et al. \cite{waring_zero-field_2020}, it was identified that the $J_\textrm{tot}$ values obtained from the fitting of spin wave frequencies at remanence were smaller than the $J_\textrm{tot}$ values obtained from the saturation field. In agreement with our conclusion of Fig.~\ref{figureVNAFMR}(a), the difference in the estimated $J_\textrm{tot}$ is minor for their 5-nm thick Co$_{20}$Fe$_{60}$B$_{20}$ layers. They report a value of $J_\textrm{tot}=-1.27 ~\textrm{mJ/m}^2$ after optimization of the Ru spacer thickness. Despite a different Fe-Co composition ratio, we believe that this number of $-1.27 ~\textrm{mJ/m}^2$ can be compared to our $-1.7\pm 0.15 ~\textrm{mJ/m}^2$. Indeed Hashimoto et al.\cite{hashimoto_fe_2006} have shown that the interlayer exchange is almost independent on the Fe-Co composition ratio in the 25-75\% and 75-25\% composition interval. 
 
Another conclusion of our study is that contrary to earlier studies \cite{wiese_antiferromagnetically_2004}, the amorphous character of our magnetic films does not prevent to achieve very strong interlayer exchange coupling, potentially as strong as in the authoritative Co/Ru(8.5 \r{A})/Co system where the interlayer coupling is typically \cite{devolder_time-resolved_2016, mckinnon_spacer_2021}  $J_\textrm{ex}=-1.5~\textrm{mJ/m}^2$.
The fact that $J_{ex}$ does not seem to decrease much with the roughness even when $\sigma_\textrm{RMS}$ and $t_{Ru}$ are of the same order of magnitude is an indication that the roughness of the spacer is conformal and that the coherence of the electron within the Ru spacer quantum well is maintained in all samples.

\section{Conclusion}
In conclusion, we have described a procedure for the measurement of the interlayer coupling and intralayer exchange stiffness in thick and thin synthetic antiferromagnets. The procedure relies on the measurement and the modeling of the frequencies of the spin wave modes confined within the thickness of the stack, and notably their dependence with the applied field. The values of the spin wave frequencies are objective and largely immune to potential imperfections in the experimental measurements, such that the procedure yields reliable estimates of the interlayer coupling and intralayer exchange stiffness. 

We have implemented our procedure on symmetric Co$_{40}$Fe$_{40}$B$_{20}$ (5-40 nm)/Ru/ Co$_{40}$Fe$_{40}$B$_{20}$ synthetic antiferromagnets in which the magnetic layers are amorphous and of controlled roughness. The exchange stiffness is found to be 16 $\pm$ 2 pJ/m, independent from the CoFeB thickness. The interlayer exchange coupling starts from -1.7 mJ/m$^2$ for the thinnest layers and it can be maintained above -1.3 mJ/m$^2$ for CoFeB layers as thick as 40 nm. Our results are compatible with the few earlier reports that carefully implemented reliable methods on comparable material systems. This comparison indicates that the amorphous character of the Co$_{40}$Fe$_{40}$B$_{20}$ layers leads to a reduced intralayer exchange stiffness but does not seem detrimental to obtain a large interlayer exchange coupling.

\section*{Acknowledgments}
A. M. acknowledges financial support from the EOBE doctoral school of Paris-Saclay University. F. M. acknowledges the French National Research Agency (ANR) under Contract No. ANR-20-CE24-0025 (MAXSAW). J. L. acknowledges the FETOPEN-01-2016-2017 [FET-Open research and innovation actions (CHIRON project: Grant Agreement ID: 801055)]. We thank Sokhna-Méry Ngom, Laurent Couraud and Ludovic Largeau for assistance in the structural characterizations.

%
%

\end{document}